\documentclass[fleqn,usenatbib]{mnras}

\usepackage{newtxtext,newtxmath}
\usepackage[T1]{fontenc}
\usepackage{ae,aecompl}

\usepackage{graphicx}	% Including figure files
\usepackage{amsmath}	% Advanced maths commands
\usepackage{amssymb}	% Extra maths symbols
\usepackage{bm}    
\usepackage{url}
\usepackage{subfig}
\usepackage{longtable}
\usepackage{booktabs,caption}
\usepackage[flushleft]{threeparttable}
\usepackage{fixltx2e}
\usepackage{adjustbox}
\usepackage{multirow}
\usepackage{ltablex}

\title[Lensing rates of GWs with detector networks]{Event rate predictions of strongly lensed gravitational waves with detector networks and more realistic templates}

\author[Yang et al.]{
Lilan Yang,$^{1,4}$
Shichao Wu,$^{6,7}$
Kai Liao,$^{1,3}$\thanks{E-mail: liaokai@whut.edu.cn}
Xuheng Ding,$^{4}$
Zhiqiang You,$^{1}$
Zhoujian Cao,$^{2}$
\and
Marek Biesiada,$^{2,5}$
Zong-Hong Zhu$^{1,2}$\thanks{E-mail: zhuzh@whu.edu.cn}
\\
\\
\\
$^{1}$School of Physics and Technology, Wuhan University, Wuhan 430072, China\\
$^{2}$Department of Astronomy, Beijing Normal University, Beijing 100875, China\\
$^{3}$School of Science, Wuhan University of Technology, Wuhan 430070, China\\
$^{4}$Kavli IPMU (WPI), UTIAS, The University of Tokyo, Kashiwa, Chiba 277-8583, Japan\\
$^{5}$National Centre for Nuclear Research, Pasteura 7, 02-093 Warsaw, Poland\\
$^{6}$Max-Planck-Institut f{\"u}r Gravitationsphysik (Albert-Einstein-Institut), D-30167 Hannover, Germany\\
$^{7}$Leibniz Universit{\"a}t Hannover, D-30167 Hannover, Germany\\
}

\date{Accepted XXX. Received YYY; in original form ZZZ}

\pubyear{2021}

\begin{document}
\label{firstpage}
\pagerange{\pageref{firstpage}--\pageref{lastpage}}
\maketitle

\begin{abstract}
Strong lensing of gravitational waves (GWs) is attracting  growing attention of the community. 
The event rates of lensed GWs by galaxies were predicted in numerous papers, which used some approximations to evaluate the GW strains detectable by a single detector. 
The joint-detection of GW signals by a network of instruments will increase the detecting ability of fainter and farther GW signals, which could increase the detection rate of the lensed GWs, especially for the 3rd generation detectors, e.g., Einstein Telescope (ET) and Cosmic Explorer (CE). 
Moreover, realistic GW templates will improve the accuracy of the prediction. 
In this work, we consider the detection of galaxy-scale lensed GW events under the 2nd, 2.5th, and 3rd generation detectors with the network scenarios and adopt the realistic templates to simulate GW signals. 
Our forecast is based on  the Monte Carlo technique which enables us to take Earth’s rotation into consideration.
We find that the overall detection rate is improved, especially for the 3rd generation detector scenarios. More precisely, it increases by $\sim$37\% adopting realistic templates, and under network detection strategy,  further increases by $\sim$58\% comparing with adoption of the realistic templates,
and we estimate that the 3rd generation GW detectors will detect hundreds lensed events per year.
The effect from the Earth's rotation is weakened in the  detector network strategy.

\end{abstract}

% Select between one and six entries from the list of approved keywords.
% Don't make up new ones.

\section{Introduction}
Since the first gravitational wave (GW) detection from binary black hole merger, the door of the new area, GW astronomy, has been opened \citep{Abbott2016}. The first detection of the GW from binary neutron star (BNS) merger accompanied by multi-band electromagnetic counterparts\citep{Abbott2017} became the landmark of multi-messenger astronomy.
The discovery of the first two neutron star-black hole events, GW200105 and GW200115, have also been announced recently,  although the electromagnetic counterparts have not been detected yet \citep{abbott2021observation}.
With the  current and upcoming observations, new paths have been opened for exploration. General relativity predicts that GW signals travel along null geodesics, like photons. 
Therefore, when a GW signal propagates near a massive foreground object, e.g., a galaxy or a cluster, it could be strongly lensed into multiple signals like the optical light is lensed into multiple images. 
During the first three observing runs (O1, O2, and O3)  Advanced LIGO and Advanced Virgo have already released two catalogs ``Gravitational-wave Transient Catalog -1 and -2” (GWTC-1 and GWTC-2) yielding 50 events in total \citep{Abbott2019, Abbott2020b}. 
Besides the main catalogs, two newly discovered neutron star-black hole events also have been released \citep{abbott2021observation}.
Moreover, searches for gravitational-wave lensing signatures have been  performed in the GWTC-1 and GWTC-2  \citep{Hannuksela2019,abbott2021search} following strategy proposed by \cite{Haris2018} and others, but no convincing evidence for lensing was found. Recently, a pair of events, GW170104 and GW170814 has been suspected to be a lensed event,  but it was not possible to confirm it convincingly \citep{Dai2020, Liu2020}.
Although no lensed GW event has been confirmed yet, future detection of  lensed GW can be expected with upcoming observations. 

Propagation of GW signal far away from the source, and its strong lensing in particular, can be described by geometric optics~\citep{Wang1996}. 
Two important parameters characterizing a strongly lensed GW event, comprise: magnifications of lensed signals and time delay between them. Unlike in the optical, we will never be able to resolve lensed GW images. The most likely imprint of GW lensing would be a detection of delayed merger signals from the same location in the sky with similar (up to magnification) temporal structure. For each lensed GW signal (i.e. for each ``image'') signal-to-noise ratios (SNRs) are augmented by a factor $\sqrt{\mu}$, where $\mu$ is the flux magnification -- a concept known from the optical observations.  
While the augmented SNR would benefit the detection, especially for farther and fainter sources~\citep{Wang1996}, the degeneracy between intrinsic chirp mass and magnification would lead to a bias in estimating the intrinsic masses of the binary star systems~\citep{Oguri2018}. Moreover, since the lensed signals arrive with delays, the detected waveforms could change due to the rotation of the Earth. This is an important practical issue discussed in \citep{ET4}, especially because of the usefulness of GW lensing time delays. Namley, the time delays between lensed GW signals can be used to test the GW propagation speed~\citep{Fan2017} and 
accurately determine the Hubble constant~\citep{Liao2017, Oguri2019}.
In the geometric optics approximation, which we adopt here, the intrinsic waveform of the lensed signal would not be distorted for a (time delay functional) minimum point and would only be  shifted by an additional phase for a saddle point~\citep{Dai2017}. 
In other cases when the lens scale is comparable to the GW  wavelength, one should take the diffraction effect into consideration, i.e., the wave optics description~\citep{Takahashi2003}.
Due to long wavelength and coherent emission of GWs,  their propagation could be different from the light.

Many works have discussed GW lensing rate using the approach  developed by \citet{Finn1993, Finn1996} to translate the 
intrinsic merger rate into the detection rate \citep{ET1, ET2, ET3, Li2018, Ng2018, ET4, ET5, Wierda2021,Wang2021,Hou2021,Mukherjee2021-ligo}.
Detection of lensed GW events enables investigating the properties of the binaries \citep{Contigiani2020, Xu2021},  constraining the cosmological parameters \citep{Hou2021}, breaking the mass-sheet degeneracy in lensing system \citep{Cremonese2021}, figuring out the binary formation channel \citep{Mukherjee2021-break}. Moreover,  \citet{Wang2021}  proposed a method to distinguish different types of lensed images. However, the key requirement necessary to pursue the above mentioned ideas is a sufficient event rate of lensed GW. With the growing number of detectors constituing the network and developing  the quality of GW templates,   
more lensed GW events are expected to be detected.

In this work, our goal is to investigate the improvement of event rate by the continuing upgrade detector network and  more realistic GW template. 
To provide some quantitative insights,  we examine the effect from above two factors, while keep other aspects of calculations as same as our previous predictions \cite{ET4},  such as mass model of lenses, the underlying source population.
First, we consider the 
networks composed of the 2th and the 3rd generation detectors (in different combinations). 
Observations based on the 2nd generation GW detectors comprising LIGO Handford, LIGO Livingston, Advanced Virgo (HLV).
Two LIGO detectors were operating alone during the fist and the second observation runs (O1 and O2). After that, when the Advanced Virgo joined in, HLV detector network began the third observation run (O3). 
Recently KAGRA has joined the HLV, and HLVK network is expected to operate at the fourth observation run (O4).
The upgrades of these detectors are still underway -- this setting we call 2.5th GW detectors, i.e., the upgrade of the Advance LIGO, Advanced Virgo and KAGRA corresponding to Advanced LIGO Plus (A+), Advanced Virgo Plus (AdV+) and KAGRA+, respectively. Some papers refer it as ``PlusNetwork'', e.g. \citet{Zhu2020}.
The PlusNetwork will come into service at the fifth observation run (O5).  
The 3rd generation of ground-based detectors like the Einstein Telescope (ET) and Cosmic Explorer (CE) is being planned. 
With continuing upgrade of the detectors, in particular with the increased sensitivity of the 3rd generation, the detection of lensed GWs is expected. 
Second, we use more realistic templates to simulate GW signal, e.g., \texttt{IMRPhenomPv3} and \texttt{IMRPhenomPv2\_NRTidalv2}
rather than an approximation.  
The detected event rates are expected to be improved by considering these two aspects. 
We perform Monte Carlo simulation to evaluate the detection rates, 
making use of the similar strategy as our previous series of papers \citep{ET1, ET2, ET3, ET4, ET5}. Hence, we will not repeat here most of the details already described in these papers.

The paper is organized as follows. We introduce the methodology of detecting a lensed GW signal emitted from compact binary coalescences (CBCs) in Section \ref{method}. Monte Carlo simulation approach to evaluate the event rate is described in Section \ref{mock}. Results and discussions are presented in Section \ref{results}. Conclusions and perspectives are given
in Section \ref{conclusion}.
Concerning the cosmological background, we assume the flat $\Lambda$CDM model with $\Omega_{\rm{M}}=0.3$ and $H_0 = \rm{70 \;km\;s^{-1}\;Mpc^{-1}}$.

\section{Methodology}\label{method}
This section introduces the methodology concerning  
three aspects: detection of GW signal, intrinsic event rates of compact object coalescence (CBC), and the lensing statistics. 

\subsection{GW signal detectability}\label{sec:gwdetect}

The detection of a GW signal from CBC is achieved by using the matched filtering method. For a single detector, the square of the signal to noise ratio (SNR) is:
\begin{equation}\label{eq:snr}
\rho^2 = 4\int_{f_\mathrm{min}}^{f_\mathrm{max}}\frac{|\tilde{h}(f)|^2}{S_n(f)}df,
\end{equation}
where the $\tilde{h}(f)$ is the signal's strain  in the frequency domain (i.e. Fourier transformed time domain signal), $S_n(f)$ is the one-sided power spectral density of the GW detector. 
The condition for registering a GW signal requires, that $\rho > \rho_0$, where $\rho_0$ is the threshold. The choice of the threshold is arbitrary and the usual choice is $\rho_0 =8.$ 
When a network consists of $i$ detectors, the optimal SNR is defined as:
\begin{equation}
\rho_\mathrm{network} = \sqrt{\rho^2_1+\rho^2_2+...+\rho^2_i}, 
\end{equation}
which means that SNRs from individual detectors are added in quadrature over all network. 

We strive to make our forecast realistic, and thus we 
employ the standard GW software \texttt{LALSuite} \citep{lalsuite}, which includes actual templates to describe waveforms from CBCs. 
For binary black hole systems (BBH) signals, we inject them into the recently improved frequency-domain precessing
quasi-circular phenomenological approximant \texttt{IMRPhenomPv3} \citep{Khan2019}.
The updated model \texttt{IMRPhenomPv3} treats the precession dynamics better, which allows for a more accurate description of double-spin precession effects, comparing to the previous single-precessing-spin model \texttt{IMRPhenomPv2} \citep{Khan2016}.
For NS related signals, we neglect tidal parameters for simplicity and make use of the model \texttt{IMRPhenomPv2\_NRTidalv2} \citep{Dietrich2019}.
The study of \citet{Zhu2020} shows that different choices of the NS's equation of state (EoS) do not change the magnitude of detection rate of the NS related systems. Hence, for NSBH systems, we also use the template \texttt{IMRPhenomPv2\_NRTidalv2} to generate signals, and we draw the intrinsic parameters of BH and NS from BBH and BNS priors, respectively.

\subsection{Intrinsic CBC rates}\label{sec:dco}

To assess the event rates of lensed GW emitted from CBCs, first we need to estimate the rates of the CBCs up to redshift $z_s$:  
 \begin{equation} \label{unlensed_rates}
{\dot N}(z_s) = \int_0^{z_s} \frac{d {\dot N}}{dz} dz,
 \end{equation}
where $\frac{d\dot{N}}{dz_{s}}$ is the merger rate of 
the CBC sources at the redshift interval $[z_{s}, z_{s}+dz_{s}]$ 
 \begin{equation}
 \frac{d\dot{N}}{dz_s}
 =4\pi\left(\frac{c}{H_{0}}\right)^3\frac{\dot{n}_{0}(z_{s})}{1+z_{s}}\frac{\tilde{r}^2(z_{s})}{E(z_{s})},
 \end{equation}
where $\dot{n}_{0}(z_{s})$ denotes the intrinsic merger rate  density at redshift $z_s$, $\tilde{r}(z_{s})$ is the dimensionless comoving distance to the source $\tilde{r}(z_{s}) = \int_0^z \frac{dz'}{E(z')}$, and $E(z_s)$ is the dimensionless (with $c/H_0$ factored out) expansion rate of the Universe at redshift $z_s$. 
We use merger rates $\dot{n}_{0}(z_{s})$ for each of the CBC systems (BBH, NSBH, BNS) as a function of cosmological redshift according to \citet{Dominik13}, which used the {\tt StarTrack} population synethesis code \citep{Belczynski2008} \footnote{We used the data available at the website \url{http://www.syntheticuniverse.org}.}. 
\citet{Dominik13} provided a suite of evolutionary models including 
the standard one and three of its modifications, i.e., Optimistic Common Envelope, delayed SN explosion and high BH kicks scenario.
The mass ratio $q=m_2/m_1$, where $m_1$ is primary mass and $m_2$ is the mass of the secondary, is a fundamental parameter for the binary system evolution.
At ZAMS {\tt StarTrack} code assumes a flat mass ratio distribution between 0 and 1, primary mass $m_1$ is assumed to follow initial mass function of Kroupa and Weidner (see \citet{Dominik13} for details). Concerning the final stages of evolution, which are relevant for detectability of GW signals, \citet{Dominik2012} reported that regardless of the  evolutionary models BH-BH systems have average mass ratio close to one, $q\sim 0.8$, similarly NS-NS have $q \sim 0.85$ only BH-NS population displays significant mass ratio $q \sim 0.2$.

In addition, to investigate the uncertainties of the chemical evolution of the Universe, \citet{Dominik13} employed two distinct scenarios for metallicity evolution, with median value of metallicity of 1.5 $Z_{\sun}$ and 0.8 $Z_{\sun}$ at $z\sim0$ labeled as high-end and low-end, respectively. 
Intrinsic merger rates predicted by different evolutionary scenarios lead to different detection rates of CBCs, (see Table 1 in \citet{ET2}). 
Among these scenarios, 
the low-end metallicity, standard scenario has been
considered in all previous works. Therefore, to make a direct comparison and understand the level of improvement, we focus only on this evolutionary scenario. 

\subsection{Lensed GW signal statistics}\label{sec:lensing}

Similar to the approach used in our previous works \cite{ET1, ET2, ET3, ET4, ET5}, we focus on galaxy-scale lensing, and assume that lensing galaxies consist of  elliptical galaxies that can be modeled as singular isothermal spheres (SIS).
Strong lensing phenomenon has a characteristic angular scale: the Einstein radius $\theta_E$, which in SIS case is $\theta_E = 4\pi(\frac{\sigma}{c})^2\frac{d_A(z_l,z_s)}{d_A(z_s)}$, where $\sigma$ is the central velocity dispersion of the lens galaxy, $d_A(z_l,z_s)$ and $d_A(z_s)$ are angular diameter distances between the lens and the source and to the source, respectively. 
Considering the source at the angular position $\beta$, if $\beta<\theta_{E}$, two strongly lensed images (brighter one $I_{+}$ and fainter one $I_{-}$) will appear on the opposite of the lens center. The corresponding angular positions of the these two images are $\theta_{\pm}=\theta_{E}\pm\beta$. 
Introducing dimensionless quantities:  $x=\frac{\theta}{\theta_{E}}$ and $y=\frac{\beta}{\theta_{E}}$,  strong lensing condition becomes $y<1$ and the lensed images appear at positions $x_{\pm}=1\pm y$ with magnifications $\mu_{\pm}=\frac{1}{y}\pm1$. 

Optical depth $\tau_{0}$ capturing the lensing probability corresponding to $y=1$ for a GW source at redshift $z_{s}$ is:
\begin{equation} \label{eq:tau0}
\small
\tau_0(z_s) = \frac{1}{4 \pi} \int_0^{z_s} dz_l  \int^{\infty}_0 d \sigma  \;
4 \pi \left( \frac{c}{H_0} \right)^3 \frac{ {\tilde r}_l^2}{E(z_l)} S_{cr}(\sigma, z_l, z_s) \frac{d n}{d \sigma}
\end{equation}
where $z_{l}$ is the redshift of lens,
and $S_{cr}$ is the cross section for lensing which reads:
\begin{equation} \label{cross_section}
S_{cr}(\sigma, z_l, z_s) = \pi \theta_E^2  = 16 \pi^3 \left( \frac{\sigma}{c} \right)^4 \left( \frac{{\tilde r}_{ls}} {{\tilde r}_{s}} \right)^2.
\end{equation}
Then, we model the velocity dispersion function (VDF) $\sigma$ in the population of lensing galaxies as a modified Schechter function $\frac{d n}{d \sigma} = n_{*} \left( \frac{\sigma}{\sigma_{*}} \right)^{\alpha} \exp{\left( - \left( \frac{\sigma}{\sigma_{*}} \right)^{\beta'} \right)} \frac{\beta'}{\Gamma (\frac{\alpha}{\beta'}) } \frac{1}{\sigma}$, with the parameters $n_{*}$,$\sigma_{*}$,${\alpha}$ and $\beta'$  taken after \citet{Choi2007}. The reason why we choose this particular velocity distribution function is that it is representative to elliptical galaxies (the population we consider to be a population of lenses). Other, most recent VDFs have been assessed on mixed population of elliptical and spiral galaxies. 
Additionally, \cite{ET2} have demonstrated that the different choice of VDFs would not affect the results  significantly.

\section{Monte Carlo strategy to evaluate the event  rates}\label{mock}

\citet{ET4} proposed a Monte Carlo simulation to assess the rates of detectable lensed GW events, taking into account the effect of Earth's rotation on event rate. This is an important ingredient. Namely, because of the rotation of the Earth the time delayed signal, even though coming from the same location in the sky as the first signal, reaches the detector from a different direction. Therefore the amplitudes of time delayed signals would be affected not only by lensing magnifications.  
We further optimize the simulation procedure, calculating the SNRs of signals through realistic CBC templates and considering detector network. We describe the main steps of these improvements and mocking procedures in this section. 

The intrinsic total number of CBC events originating in the redshift interval  $[z_{s}, z_{s}+dz_{s}]$ is determined by {\tt StarTrack} population under low-end metallicity, standard scenario.
In one realization, we sample the lensed events according to  the probability $\tau_0(z_s)$ (given by Eq.~(\ref{eq:tau0})). 
We perform this strategy for CBC events up to redshift z$\sim$18 and obtain $\tau_0(z_s)$ fraction of lensed CBCs at redshift $z_s$ randomly. Since the optical depth for lensing 
is a small number, this strategy avoids 
wasting computational time on the non-lensed events.

Once a CBC event satisfies the strong lensing condition, two lensed signals $I_{-}$ and $I_{+}$ will arrive at the detector at different times $t_-$ and $t_+$.
We evaluate the intrinsic SNR of them
$\rho_{intr.\pm}$  according to Eq.~(\ref{eq:snr}), where
the GW strain signal $\tilde{h}(f)$ is calculated
via realistic templates in the geocentric frame.
The templates depend on source parameters, which include eight intrinsic parameters describing the
quasi-circular CBC systems (i.e., primary and secondary masses $m_{1}$ and $m_{2}$, spin parameters, spin angles) and seven 
external parameters (i.e., luminosity distance $d_L$, inclination angle $\theta_{JN}$, polarization angle $\psi$, phase of coalescence $\phi_c$, right ascension RA and declination DEC, and time of coalescence $t_{c}$ in geocentric frame~\citep{Ashton2019}).
The prior distributions of the source parameters for BBH systems and BNS systems are presented in Table~\ref{tab:bbh} similarly as in \citet{Abbott2020a}.

For BBH systems, we assume that masses $m_1$, $m_2$ are distributed according to a power-law with the index of $\alpha=-2.3$, mass range 5-50 $M_{\sun}$, and spins aligned or anti-aligned with uniformly distributed magnitudes between (0, 0.99). We only consider their dimensionless spin magnitudes $a_{1,2}$, and other four spin angles are set to zero.
For BNS systems, we adopt the Gaussian distribution of $m_1$, $m_2$ with mean $\mu_m =1.33$ and standard deviation $\sigma_m = 0.09$, 
based on BNS properties found in Milky Way, although this distribution has been challenged \citep{Farrow2019}.
The spins aligned or anti-aligned with uniformly distributed magnitudes between (0, 0.05).
For NSBH systems, the distributions of masses and spins 
are the combination of BNS and BBH populations.
The detection rate depends on the CBC's mass function, which is still not well established. 
Thus as a supplementary material, we also present the results obtained with different choice of mass function parameters. Namely, for BBH we adopt the power-law index of $\alpha=-1$, and mass range $5-85 \;M_{\sun}$, the upper boundary mass value referring to GW190521 \citep{GW190521}.
For BNS, we set the uniform distribution between $0.8-2.3 \; M_{\sun}$, and the combination of BNS and BBH populations for NSBH.
The source parameters of two lensed signals $I_-$ and $I_+$ are identical, but note that $\rho_{intr._-}$ and $\rho_{intr._+}$ are differently registered in the geocentric frame due to the time delay between them and the fact that during this period the Earth rotates \citep{ET4}.

%Table1 
\begin{table}
\centering
\caption{Priors for BBH and BNS. 
The intrinsic variables contain two black hole masses $m_{1,2}$ and their dimensionless spin magnitudes $a_{1,2}$. The extrinsic parameters are the luminosity distance $d_L$, the right ascension RA and declination DEC, the inclination angle $\theta_{JN}$ between the observer's line of sight and the total angular momentum, the polarization angle $\psi$, and the phase at coalescence $\phi_c$. We randomly generate the coalescence time $t$ in geocentric frame.}
\begin{tabular}{lcccc}
\hline
\hline
parameters & unit & shape & min & max \\
\hline
$m_{1,2}$ (BBH) & $M_{\sun}$ & Power-law, $\alpha=-2.3$ & 5 &  50 \\
$m_{1,2}$ (BNS)& $M_{\sun}$ & Gaussian, \textit{N}(1.33, 0.09) & - & - \\

$a_{1,2}$ (BBH) & 1 & Uniform & -0.99 & 0.99 \\
$a_{1,2}$ (BNS) & 1 & Uniform & -0.05 & 0.05 \\
$d_{L}$ & Mpc & UniformSourceFrame  &  100 & 198657.3 \\
RA & rad. & Uniform & 0 & $2\pi$ \\
DEC & rad. & Cos & $-\pi/2$  & $\pi/2$ \\
$\theta_{JN}$ & rad. & Sin & 0 & $\pi$ \\
$\psi$ & rad. & Uniform & 0 & $\pi$ \\
$\phi_c$ &  rad. & Uniform & 0 & $2\pi$\\
$t$ & s & - & - & - \\
\hline
\label{tab:bbh}
\end{tabular}
\end{table}

The intrinsic SNRs  of two lensed signals, denoted by  
$\rho_{intr._\pm}$ will be modified by lensing magnification to
$\rho_{\pm}=\rho_{intr._\pm}\sqrt{\mu_{\pm}}=\rho_{intr._\pm}\sqrt{\frac{1}{y}\pm1}$, where the dimensionless quantity $y^2$ follows a uniform  distribution ${\cal U}([0, 1])$.

Detection of a lensed pair of signals requires that both $\rho_{\pm}$ exceed the detection threshold. 
Here, we consider network scenarios from the 2nd to 3rd generation of GW detectors.
For the 2nd generation detectors, HLV and HLVK, we assume a CBC to be detected if its has $\rho>4$ in at least two detectors and a network SNR should be $\rho_{network}>12$ \citep{Abbott2020a}.
For 2.5th generation detectors, aka PlusNetwork, we keep the same detection criteria as for the 2nd generation.
For the 3rd generation detectors, we consider two conditions,  ET alone and the combination of ET and CE.
We require both a single and a network have $\rho, \rho_{network} > 8$.
Network detection lowers the threshold of signal at a single detector, hence a signal is easier to be detected by a network detectors than in one detector alone. 

We preformed more than $10^6$ realizations for all CBC systems and accumulated the events that meet the requirements above to obtain the event rates.
In our simulation,  the sensitive curves expressed as the amplitude spectral densities (ASD)
are adopted from the public data
for the 2nd \footnote{\url{https://dcc.ligo.org/LIGO-P1200087-v42/public}}
and 2.5th \footnote{\url{https://gwdoc.icrr.u-tokyo.ac.jp/cgi-bin/DocDB/ShowDocument?docid=9537}}
generation detectors. For the 3rd generation detectors, we have selected ``ET-D" (xylophone design) and ``CE1" for ET \footnote{\url{http://www.et-gw.eu/index.php/etsensitivities}} and CE \footnote{\url{https://dcc.cosmicexplorer.org/cgi-bin/DocDB/ShowDocument?docid=T2000017}}
from the official websites, respectively.

\section{Results and Discussion}\label{results}
Table~\ref{tab:lensed} shows the expected yearly detection rates of galaxy-scale lensed GWs based on the standard model of CBC formation in low-end metallicity scenario calculated up to the 3rd generation of GW detectors network, with mass distribution of BBH and BNS following power-law with $\alpha=-2.3$, mass range 5-50 $M_{\sun}$ and Gaussian distribution N(1.22, 0.99), respectively.
One can see that the lensed GW signals have negligible chance to be detected under the design sensitivity of both the 2nd and 2.5th generation of detectors.
Fortunately, when GW detectors are upgraded to the 3rd generation,  the total number of the lensed GWs significantly increases to $>200$ for ET and $>350$ for combined ET+CE network. 
The results obtained with alternative GW sources mass distribution are presented in Table \ref{tab:lensed-diffmass}. 
We note that the predictions for both BNS and NSBH populations do not differ much.  
For BBH events, however, the number of lensed BBH events increases for PlusNetwork, ET and ET+CE as expected, 
see Appendix \ref{app} for details.

Total number of lensed GW events detectable by ET increases by 37\% with respect to the forecasts obtained without using templates to generate GW signals as presented in \citet{ET4} (see Xylophone design in their Table 1). 
Especially, the detection rate of BNS is increased by more than 50\%. 
The increase of the event rate is associated with the fact that the prediction performed in this work is more realistic,  because of adopting reliable GW templates.
Moreover, to mock GW signals as authentic as possible, we assigned masses of compact objects via a distribution rather than a fixed value and the spin parameters are also considered.  
As expected the network detection will boost the event rate. It is because the network setting decreases the detection threshold by a factor about $\sqrt{i}$, where $i$ is the number of detectors. Taking 3rd generation detector network ET and CE as an example, with additional detector CE, the rates of lensed GWs (both $I_-$ and $I_+$ detected) are 3.88, 7.13 and 342.93 for BNS, NSBH, and BBH, which improves capabilities of the ET alone by  2.5, 1.1 and 0.6 times, respectively. 
The detector network benefits not only from decrease of detection threshold, but also from the reduction of the effect of the Earth's rotation. Due to the Earth's rotation, the SNRs for two lensed images $I_-$ and $I_+$ are not only affected by lens magnification, hence the detection of the fainter $I_-$ image no longer guarantees the detection of the brighter one $I_+$. Interplay between time delay and Earth's rotation complicates the issue by changing direction to the signal with respect to the detector. But multiple detectors spread on the Earth counterbalance this effect. 
For example, for the BBH systems in the 3rd generation network case, the difference between $I_-$ and $(I_-,I_+)$ detection rate decreases from 3\% with a single ET detector to 0.5\% with the addition of CE detector.

We conservatively discussed galaxy-scale lensing assuming the SIS model (producing two images), for the purpose of discussing the relative increase of event rate considering the detector networks and realistic GW template comparing to our previous works.
The quadruple (four-images) %and cusp (three-images) 
lensing systems have been ignored, because according to estimates of \citep{Li2018} they comprise a fraction of about  6\% of all strong lensing systems detectable by the ET. Galaxy clusters are also an important population of strong lenses, especially regarding high magnification lensing events. 
\cite{Robertson2020} reported that they found half of all high magnification lensing ($\mu>10$) is due to galaxy groups and clusters while the other half is due to galaxies.
The numbers of GW lensing event will certainly increase with consideration of cluster-scale lensing, 
although the fraction of observed lensed events at cluster-scale is much rarer than the rates at galaxy-scale \citep{abbott2021search}. Again, we do not discuss it in more detailed way for the sake of comparison with previous forecasts.

The models to describe CBC merger rates as a function of redshift are closely related to the redshift evolution of star formation rate (SFR), whose most popular form is that of \citep{Madau2014}. 
Our prediction is based on merger rate predicted by {\tt StarTrack} code, which is based on the SFR model due to \cite{Strolger2004}. Hence, one may worry about the reliability of {\tt StarTrack} model.
On the one hand,
we present the expected yearly detection rate of non-lensed GW in Table~\ref{tab:non-lensed}.
Comparison with the rates of lensed events indicates the lensing probability, and we find that under  most detection scenarios, lensing probability is about $\sim$ 0.1\%, in accordance with results reported in many previous works, such as earlier analytic prediction \citep{Turner1984}, simulation of the fraction of lensed quasar \citep{Pindor2003}, 
and GW lensing rate prediction \citep{Wang2021}. 
However, the lensing rates of BNS and NSBH under the 2nd and 2.5th generation detectors scenarios are much smaller than the overall lensing probability.
Two factors explaining this are: first, the distribution of BNS and NSBH populations mainly located at lower
redshifts resulting in smaller lensing probability than the overall
value.
Second, neglecting high magnification contributed from cluster lensing, which leads to the underestimation \citep{Smith2018}. 
In other words, the overall 
lensing probability is the integrated optical depth $\tau(>\mu) \sim \mu^{-2}$ to lens magnification of $\mu > 2$, which is the threshold of image pair formation in SIS model. 
In order to be detectable before ET+CE era, such sources need to be magnified at least by factors of about  $\mu \sim 100$.  
Neglecting cluster lensing, we are essentially ignoring 
the high magnification tail, where the lensed BNS and NSBH population in pre-3G era could come from (if detected).

We also note that the predicted detection rate of CBC signals for the 2nd generation is at the same level as the up-to-date reported detected signals in GWTC-2.
While the HLV network has not reached its design sensitivity,
it has discovered 39 candidate events during $\sim$26 weeks ($\sim$1.5 per week).
We predict that this network, with design sensitivity, can detect $\sim$141 candidates during one year ($\sim$2.7 per week). 
It can be expected that during the 4th observing run, HLVK network would detect $\sim$5 events per week. 
On the other hand, the GW lensing rate and CBC's merger rate history can be constrained by the amplitude of the stochastic gravitational wave background (SGWB). \cite{Mukherjee2021-ligo} demonstrated a close relation between the GW lensing rate and the expected level of SGWB as function of magnification faction for different merger rates of CBC sources. 
Also, the BBH merger rate as a function of redshift can be constrained by the SGWB and BBH detection as demonstrated in \cite{Callister2020, Abbott2021upper}.
The posterior constraints on BBH merger rate history has been presented, especially at high redshift (see Figure 6 in \cite{Abbott2021upper}), e.g., upper limits is about $10^3$/Gpc$^{3}$/yr at z$>$2. The merger rate evolution of BBH adopted in this work (see Figure 3 in \citet{Dominik13}) is consistent with the limitation. 
Moreover, since the first detections of GWs and after HLVK network became fully operative, {\tt StarTrack} code predictions have been confronted with empirical data \cite{Belczynski2020}. Most recently \cite{Belczynski2021} discussed the  {\tt StarTrack} predictions of SGWB due to unresolved CBC for HLVK and ET. Even though some non-standard  original scenarios might be incompatible with the current data, yet the low-metallicity standard scenario adopted here is compatible with the HLVK data. 
Furthermore, we checked the mass ratio distribution of detected binary systems in our simulation, and the result shows that systems with mass ratio $q \sim$ 0.6 dominate the detected population, which is slightly smaller than
the empirical results released by LIGO/Virgo/KAGRA collaboration. In fact, we also note that the latest results \citep{2021LVK} support the systems with substantially smaller mass ratio. In summary, the above  aspects confirm the accuracy of the {\tt StarTrack} model.

\begin{figure}
    \centering
    \includegraphics[trim=1mm 1mm 1mm 1mm, clip, width=80mm]{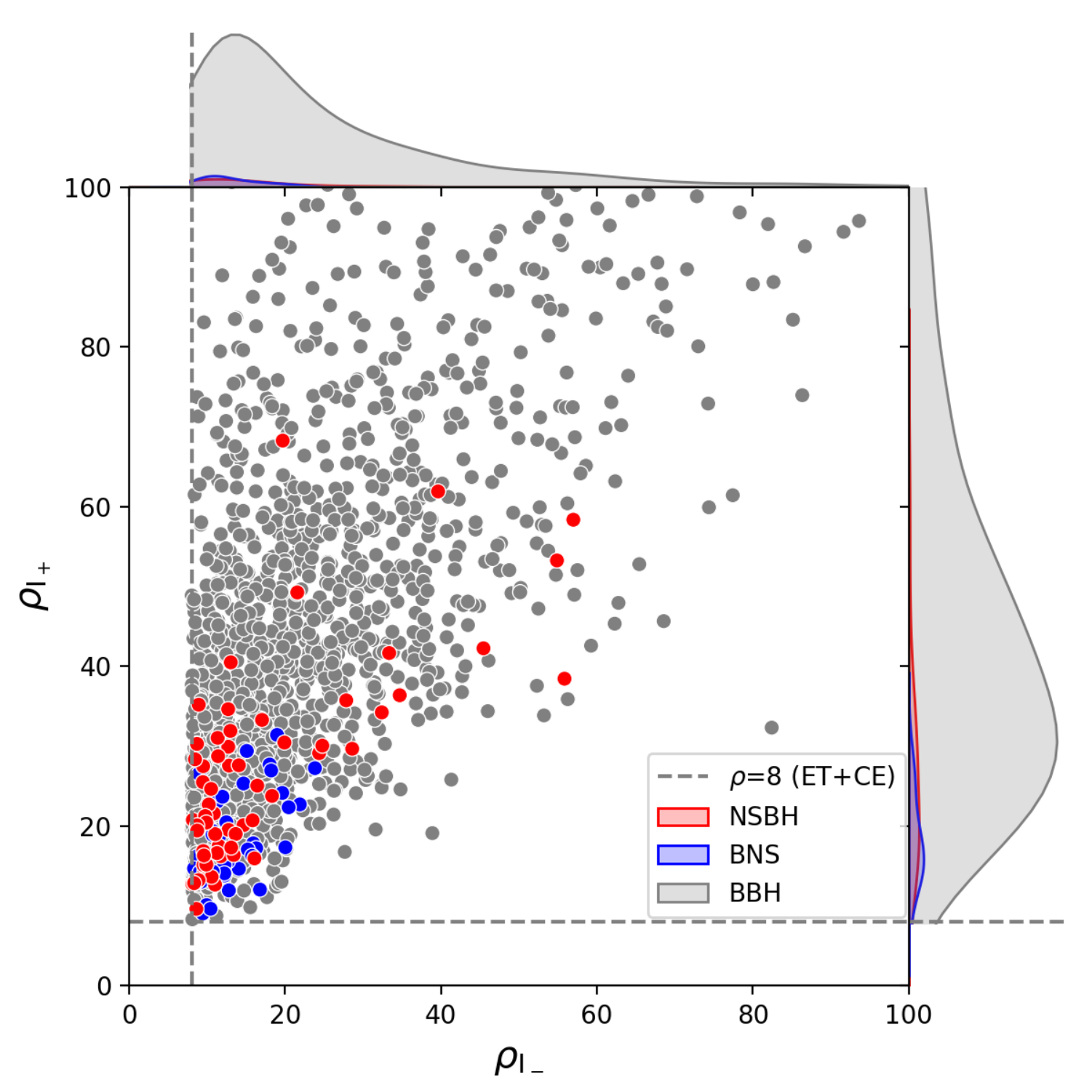}
    \caption{SNR comparison of lensing pairs, $I_-$ and $I_+$, under 3rd generation network detection scenario. Grey, blue and red colors represent the lensing pairs from BBH, BNS, and NSBH, respectively. 
    The panels on the top and right show the SNR distribution of $I_-$ and $I+$, respectively.
    The dashed line shows the detector's detection threshold. }
    \label{fig:rhorho}
\end{figure}

Based on our prediction, 200-400 lensed GW events would be registered by future 3rd generation detectors.
Obviously, not all lensed events predicted in this work can be identified, because it is challenging to discriminate whether a pair of GW signals are lensed or not.
Only those events with highly overlapped parameter space can be statistically identified as pairs. We plot the distributions of the SNRs of lensed CBCs in Figure \ref{fig:rhorho}.
There are few pairs have $\rho_{I_+}<\rho_{I_-}$, which is due to the Earth's rotation resulting in $\rho_{intr._-}>\rho_{intr._+}$.
It is expected that a pair of events, where both lensed signals have higher SNRs is easier to be identified. 
For a pure GW signal without electromagnetic counterparts, it is expected that lensed BBHs are the primary targets since they dominate the scatter plot part in Figure \ref{fig:rhorho}, with SNR constrating on 20-30. A quantitative analysis based on the method proposed by \citet{Haris2018} is worth conducting in the future.

%lensed GW
\begin{table*}
\centering
\caption{Predictions of yearly lensed GW detection rates for which only $I_-$ image or both $I_-$ and $I_{+}$ images are magnified above the threshold
under the 2nd, 2.5th and 3rd generation detectors scenarios.  Results are shown for the standard model of CBC formation in  low-end metallicity scenario.
The mass distribution of BBH follows power-law with $\alpha=-2.3$ in the mass range 5-50 $M_{\sun}$, BNS follows Gaussian distribution N(1.22, 0.99), and NSBH follows the combination of two above populations.}

\begin{tabular}{lcccccccccccr}
\hline
\hline
CBC type & \multicolumn{2}{c}{HLV} & \multicolumn{2}{c}{HLVK} & \multicolumn{2}{c}{PlusNetwork} & \multicolumn{2}{c}{ET} & \multicolumn{2}{c}{ET+CE} \\

  &  I- & I-,I+  & I- & I-,I+ & I- & I-,I+ & I- & I-,I+ & I- & I-,+ \\
\hline
BNS & $<10^{-6}$ & $<10^{-6}$ & $<10^{-6}$ &$<10^{-6}$  & $<10^{-6}$ & $<10^{-6}$ & 1.55 & 1.10 & 4.06 & 3.88\\
\hline
NSBH & $<10^{-6}$ & $<10^{-6}$ & $<10^{-6}$ & $<10^{-6}$  & $<10^{-6}$ & $<10^{-6}$ & 3.78 & 3.40& 7.21 & 7.13\\
\hline
BBH & 0.33 & 0.24 & 0.31& 0.29 & 0.70 & 0.62 & 226.19& 219.22 & 344.95 & 342.93\\
\hline
Total & 0.33 &0.24&0.31&0.29&0.70&0.62 & 231.52&223.72 & 356.22 & 353.94\\
\hline
\label{tab:lensed}
\end{tabular}

\end{table*}

\begin{table*}
\centering
\caption{Predictions of yearly GW detection rates (non-lensed signals) under the 2nd, 2.5th and 3rd generation detectors scenarios. Results are shown for the standard model of CBC formation in low-end metallicity scenario.}
\begin{tabular}{lcccccr}
\hline
\hline
CBC type & HLV & HLVK & PlusNetwork & ET & ET+CE \\
\hline
BNS & 4.92 &39.28 & 43.18 & 4851.70 & 10737.11 \\
\hline
NSBH  & 5.11 & 39.33 & 211.25 & 14888.29 & 26101.67\\
\hline
BBH & 130.94 & 164.34 & 1104.75 & 261993.09 & 337973.94\\
\hline
Total & 140.97 & 242.95 & 1747.80 & 281733.08 & 374812.72\\
\hline
\label{tab:non-lensed}
\end{tabular}

\end{table*}

\section{conclusion}\label{conclusion}
In this work, we used Monte Carlo simulation technique to 
estimate the galaxy-scale strong lensing rates of GWs for the detector networks of the 2nd and 3rd generation.
We have used the {\tt StarTrack} population synthesis code to estimate the intrinsic merger rate of CBCs.
To make our prediction realistic, 
we adopted the reliable templates, e.g., \texttt{IMRPhenomPv3} for BBH,  and \texttt{IMRPhenomPv2\_NRTidalv2} for BNS and NSBH, and we considered the effect of the Earth's rotation on the detection of time delayed signals. 
We assumed that mass distribution of BBH and BNS followed power-law with $\alpha=-2.3$ in mass range 5-50 $M_{\sun}$ and Gaussian distribution N(1.22, 0.99), respectively, and we found that,

\begin{itemize}
\item{
1. Total yearly detection rates of lensed GW signals from CBCs are respectively: 0.24, 0.29, 0.62, 223.72 and 353.94 for the 2nd generation detector network, i.e. HLV and HLVK, 2.5th generation ``PlusNetwork'' and the 3rd generation detectors, i.e. ET and ET+CE. 
For BNS and NSBH systems, 
the detection probability is extremely low
for the 2nd and 2.5th generations of detector, 
while a few lensed signals from such systems can be detected yearly by the 3rd generation detectors. 
The total event rate is dominated by BBH systems which 
would be primary candidates for searching lensed pairs of GW signals.
}

\item{ 
2. 200-400 lensed GW events would be detected by the 3rd generation detectors.
The rate increases by 37\%  with respect to previous estimates due to application of realistic signal templates. Moreover, the joint detection by a couple of 3rd generation detectors, i.e. ET+CE, further improves the event rate by about 58\% versus
a single detector, e.g. ET.  
We also observed that the multiple detectors alleviate the effect from the Earth's rotation on lensed GW detection rate.}
\end{itemize}

%lensed GW with alpha=-1 for bhbh and flat 0.8-2.3msun for nsns
\begin{table*}
\centering
\caption{The same as Table \ref{tab:lensed}, but for BBH modeled with power-law mass distribution with $\alpha$=-1 in the mass range 5-85 $M_{\sun}$, BNS modeled with flat mass distribution in the mass range 0.8-2.3 $M_{\sun}$, and NSBH modeled as a combination of the above two.}
\begin{tabular}{lcccccccccccr}
\hline
\hline
CBC type & \multicolumn{2}{c}{HLV} & \multicolumn{2}{c}{HLVK} & \multicolumn{2}{c}{PlusNetwork} & \multicolumn{2}{c}{ET} & \multicolumn{2}{c}{ET+CE} \\
&  I- & I-,I+  & I- & I-,I+ & I- & I-,I+ & I- & I-,I+ & I- & I-,+ \\
\hline
%BNS & 0 & 0 & 0&0  & 0 & 0 & 1.55 & 1.10 & 4.06 & 3.88\\
BNS & $<10^{-6}$ & $<10^{-6}$ & $<10^{-6}$ & $<10^{-6}$  & $<10^{-6}$ & $<10^{-6}$  & 1.78 & 1.22 & 4.54&  4.32\\
\hline
%NSBH & 0 & 0 & 0&0 & 0 & 0 & 3.78 & 3.40& 7.21 & 7.13\\
NSBH & $<10^{-6}$ & $<10^{-6}$ & $<10^{-6}$ &$<10^{-6}$  & $<10^{-6}$ & $<10^{-6}$ & 4.84 & 4.53 & 7.30 & 7.23\\
\hline

BBH & 0.39 & 0.17 & 0.73 & 0.55 & 3.14 & 2.74 & 333.51 & 329.83& 439.97 & 437.48 \\
%BBH & 0.33 & 0.24 & 0.31& 0.29 & 0.70 & 0.62 & 226.19& 219.22 & 344.95 & 342.93\\
\hline
%Total & 0.33 &0.24&0.31&0.29&0.70&0.62 & 231.52&223.72 & 356.22 & 353.94\\
Total & 0.39 & 0.17 & 0.73 & 0.55 & 3.14 & 2.74 & 340.13 & 335.58 & 451.81 & 449.03  \\
\hline

\label{tab:lensed-diffmass}
\end{tabular}

\end{table*}

\section*{Acknowledgments}
We would like to thank the referee for careful reading of the manuscript and stimulating comments, which allowed to improve the paper substantially.
This work was supported by the National Natural Science Foundation of China under Grants Nos. 11973034, 11633001, 11920101003 and 12021003, and the Strategic Priority Research Program of the Chinese Academy of Sciences, Grant No. XDB23000000. M.B. was supported by the Key Foreign Expert Program for the Central Universities No. X2018002.

\section*{Data Availability} 

The data underlying this article are available in the article and in its online supplementary material.

%%%%%%%%%%%%%%%%%%%%%%%%%%%%%%%%%%%%%%%%%%%%%%%%%%

%%%%%%%%%%%%%%%%%%%% REFERENCES %%%%%%%%%%%%%%%%%%

% The best way to enter references is to use BibTeX:

\bibliographystyle{mnras}
\bibliography{reference} 

\begin{thebibliography}{}
\makeatletter
\relax
\def\mn@urlcharsother{\let\do\@makeother \do\$\do\&\do\#\do\^\do\_\do\%\do\~}
\def\mn@doi{\begingroup\mn@urlcharsother \@ifnextchar [ {\mn@doi@}
  {\mn@doi@[]}}
\def\mn@doi@[#1]#2{\def\@tempa{#1}\ifx\@tempa\@empty \href
  {http://dx.doi.org/#2} {doi:#2}\else \href {http://dx.doi.org/#2} {#1}\fi
  \endgroup}
\def\mn@eprint#1#2{\mn@eprint@#1:#2::\@nil}
\def\mn@eprint@arXiv#1{\href {http://arxiv.org/abs/#1} {{\tt arXiv:#1}}}
\def\mn@eprint@dblp#1{\href {http://dblp.uni-trier.de/rec/bibtex/#1.xml}
  {dblp:#1}}
\def\mn@eprint@#1:#2:#3:#4\@nil{\def\@tempa {#1}\def\@tempb {#2}\def\@tempc
  {#3}\ifx \@tempc \@empty \let \@tempc \@tempb \let \@tempb \@tempa \fi \ifx
  \@tempb \@empty \def\@tempb {arXiv}\fi \@ifundefined
  {mn@eprint@\@tempb}{\@tempb:\@tempc}{\expandafter \expandafter \csname
  mn@eprint@\@tempb\endcsname \expandafter{\@tempc}}}

\bibitem[\protect\citeauthoryear{{Abbott} et~al.,}{{Abbott}
  et~al.}{2016}]{Abbott2016}
{Abbott} B.~P.,  et~al., 2016, \mn@doi [\prl] {10.1103/PhysRevLett.116.061102},
  \href {https://ui.adsabs.harvard.edu/abs/2016PhRvL.116f1102A} {116, 061102}

\bibitem[\protect\citeauthoryear{Abbott et~al.,}{Abbott
  et~al.}{2017}]{Abbott2017}
Abbott B.~P.,  et~al., 2017, \mn@doi [Phys. Rev. Lett.]
  {10.1103/PhysRevLett.119.161101}, 119, 161101

\bibitem[\protect\citeauthoryear{{Abbott} et~al.,}{{Abbott}
  et~al.}{2019}]{Abbott2019}
{Abbott} B.~P.,  et~al., 2019, \mn@doi [Physical Review X]
  {10.1103/PhysRevX.9.031040}, \href
  {https://ui.adsabs.harvard.edu/abs/2019PhRvX...9c1040A} {9, 031040}

\bibitem[\protect\citeauthoryear{{Abbott} et~al.,}{{Abbott}
  et~al.}{2020a}]{Abbott2020b}
{Abbott} R.,  et~al., 2020a, arXiv e-prints, \href
  {https://ui.adsabs.harvard.edu/abs/2020arXiv201014527A} {p. arXiv:2010.14527}

\bibitem[\protect\citeauthoryear{{Abbott} et~al.,}{{Abbott}
  et~al.}{2020b}]{Abbott2020a}
{Abbott} B.~P.,  et~al., 2020b, \mn@doi [Living Reviews in Relativity]
  {10.1007/s41114-020-00026-9}, \href
  {https://ui.adsabs.harvard.edu/abs/2020LRR....23....3A} {23, 3}

\bibitem[\protect\citeauthoryear{{Abbott}, {Abbott}, {Abraham}  \&
  {Acernese}}{{Abbott} et~al.}{2020c}]{GW190521}
{Abbott} R.,  {Abbott} T.~D.,  {Abraham} S.,   {Acernese} 2020c, \mn@doi [\prl]
  {10.1103/PhysRevLett.125.101102}, \href
  {https://ui.adsabs.harvard.edu/abs/2020PhRvL.125j1102A} {125, 101102}

\bibitem[\protect\citeauthoryear{Abbott et~al.,}{Abbott
  et~al.}{2021a}]{abbott2021search}
Abbott R.,  et~al., 2021a, arXiv preprint arXiv:2105.06384

\bibitem[\protect\citeauthoryear{{Abbott}, {Abbott}, {Abraham}  \&
  {Acernese}}{{Abbott} et~al.}{2021b}]{Abbott2021upper}
{Abbott} R.,  {Abbott} T.~D.,  {Abraham} S.,   {Acernese} F.,  2021b, \mn@doi
  [\prd] {10.1103/PhysRevD.104.022004}, \href
  {https://ui.adsabs.harvard.edu/abs/2021PhRvD.104b2004A} {104, 022004}

\bibitem[\protect\citeauthoryear{Abbott et~al.,}{Abbott
  et~al.}{2021c}]{abbott2021observation}
Abbott R.,  et~al., 2021c, The Astrophysical Journal Letters, 915, L5

\bibitem[\protect\citeauthoryear{{Ashton} et~al.,}{{Ashton}
  et~al.}{2019}]{Ashton2019}
{Ashton} G.,  et~al., 2019, \mn@doi [\apjs] {10.3847/1538-4365/ab06fc}, \href
  {https://ui.adsabs.harvard.edu/abs/2019ApJS..241...27A} {241, 27}

\bibitem[\protect\citeauthoryear{{Belczynski, K.} et~al.,}{{Belczynski, K.}
  et~al.}{2020}]{Belczynski2020}
{Belczynski, K.} et~al., 2020, \mn@doi [A\&A] {10.1051/0004-6361/201936528},
  636, A104

\bibitem[\protect\citeauthoryear{{Belczynski}, {Kalogera}, {Rasio}, {Taam},
  {Zezas}, {Bulik}, {Maccarone}  \& {Ivanova}}{{Belczynski}
  et~al.}{2008}]{Belczynski2008}
{Belczynski} K.,  {Kalogera} V.,  {Rasio} F.~A.,  {Taam} R.~E.,  {Zezas} A.,
  {Bulik} T.,  {Maccarone} T.~J.,   {Ivanova} N.,  2008, \mn@doi [\apjs]
  {10.1086/521026}, \href
  {https://ui.adsabs.harvard.edu/abs/2008ApJS..174..223B} {174, 223}

\bibitem[\protect\citeauthoryear{{Biesiada}, {Ding}, {Pi{\'o}rkowska}  \&
  {Zhu}}{{Biesiada} et~al.}{2014}]{ET2}
{Biesiada} M.,  {Ding} X.,  {Pi{\'o}rkowska} A.,   {Zhu} Z.-H.,  2014, \mn@doi
  [\jcap] {10.1088/1475-7516/2014/10/080}, \href
  {http://adsabs.harvard.edu/abs/2014JCAP...10..080B} {10, 080}

\bibitem[\protect\citeauthoryear{{Callister}, {Fishbach}, {Holz}  \&
  {Farr}}{{Callister} et~al.}{2020}]{Callister2020}
{Callister} T.,  {Fishbach} M.,  {Holz} D.~E.,   {Farr} W.~M.,  2020, \mn@doi
  [\apjl] {10.3847/2041-8213/ab9743}, \href
  {https://ui.adsabs.harvard.edu/abs/2020ApJ...896L..32C} {896, L32}

\bibitem[\protect\citeauthoryear{{Choi}, {Park}  \& {Vogeley}}{{Choi}
  et~al.}{2007}]{Choi2007}
{Choi} Y.-Y.,  {Park} C.,   {Vogeley} M.~S.,  2007, \mn@doi [\apj]
  {10.1086/511060}, \href {http://adsabs.harvard.edu/abs/2007ApJ...658..884C}
  {658, 884}

\bibitem[\protect\citeauthoryear{{Contigiani}}{{Contigiani}}{2020}]{Contigiani2020}
{Contigiani} O.,  2020, \mn@doi [\mnras] {10.1093/mnras/staa026}, \href
  {https://ui.adsabs.harvard.edu/abs/2020MNRAS.492.3359C} {492, 3359}

\bibitem[\protect\citeauthoryear{{Cremonese}, {Ezquiaga}  \&
  {Salzano}}{{Cremonese} et~al.}{2021}]{Cremonese2021}
{Cremonese} P.,  {Ezquiaga} J.~M.,   {Salzano} V.,  2021, \mn@doi [\prd]
  {10.1103/PhysRevD.104.023503}, \href
  {https://ui.adsabs.harvard.edu/abs/2021PhRvD.104b3503C} {104, 023503}

\bibitem[\protect\citeauthoryear{{Dai} \& {Venumadhav}}{{Dai} \&
  {Venumadhav}}{2017}]{Dai2017}
{Dai} L.,  {Venumadhav} T.,  2017, arXiv e-prints, \href
  {https://ui.adsabs.harvard.edu/abs/2017arXiv170204724D} {p. arXiv:1702.04724}

\bibitem[\protect\citeauthoryear{{Dai}, {Zackay}, {Venumadhav}, {Roulet}  \&
  {Zaldarriaga}}{{Dai} et~al.}{2020}]{Dai2020}
{Dai} L.,  {Zackay} B.,  {Venumadhav} T.,  {Roulet} J.,   {Zaldarriaga} M.,
  2020, arXiv e-prints, \href
  {https://ui.adsabs.harvard.edu/abs/2020arXiv200712709D} {p. arXiv:2007.12709}

\bibitem[\protect\citeauthoryear{{Dietrich}, {Samajdar}, {Khan},
  {Johnson-McDaniel}, {Dudi}  \& {Tichy}}{{Dietrich}
  et~al.}{2019}]{Dietrich2019}
{Dietrich} T.,  {Samajdar} A.,  {Khan} S.,  {Johnson-McDaniel} N.~K.,  {Dudi}
  R.,   {Tichy} W.,  2019, \mn@doi [\prd] {10.1103/PhysRevD.100.044003}, \href
  {https://ui.adsabs.harvard.edu/abs/2019PhRvD.100d4003D} {100, 044003}

\bibitem[\protect\citeauthoryear{{Ding}, {Biesiada}  \& {Zhu}}{{Ding}
  et~al.}{2015}]{ET3}
{Ding} X.,  {Biesiada} M.,   {Zhu} Z.-H.,  2015, \mn@doi [\jcap]
  {10.1088/1475-7516/2015/12/006}, \href
  {http://adsabs.harvard.edu/abs/2015JCAP...12..006D} {12, 006}

\bibitem[\protect\citeauthoryear{{Dominik}, {Belczynski}, {Fryer}, {Holz},
  {Berti}, {Bulik}, {Mandel}  \& {O'Shaughnessy}}{{Dominik}
  et~al.}{2012}]{Dominik2012}
{Dominik} M.,  {Belczynski} K.,  {Fryer} C.,  {Holz} D.~E.,  {Berti} E.,
  {Bulik} T.,  {Mandel} I.,   {O'Shaughnessy} R.,  2012, \mn@doi [\apj]
  {10.1088/0004-637X/759/1/52}, \href
  {https://ui.adsabs.harvard.edu/abs/2012ApJ...759...52D} {759, 52}

\bibitem[\protect\citeauthoryear{{Dominik}, {Belczynski}, {Fryer}, {Holz},
  {Berti}, {Bulik}, {Mandel}  \& {O'Shaughnessy}}{{Dominik}
  et~al.}{2013}]{Dominik13}
{Dominik} M.,  {Belczynski} K.,  {Fryer} C.,  {Holz} D.~E.,  {Berti} E.,
  {Bulik} T.,  {Mandel} I.,   {O'Shaughnessy} R.,  2013, \mn@doi [\apj]
  {10.1088/0004-637X/779/1/72}, \href
  {http://adsabs.harvard.edu/abs/2013ApJ...779...72D} {779, 72}

\bibitem[\protect\citeauthoryear{Fan, Liao, Biesiada, Pi\'orkowska-Kurpas  \&
  Zhu}{Fan et~al.}{2017}]{Fan2017}
Fan X.-L.,  Liao K.,  Biesiada M.,  Pi\'orkowska-Kurpas A.,   Zhu Z.-H.,  2017,
  \mn@doi [Phys. Rev. Lett.] {10.1103/PhysRevLett.118.091102}, 118, 091102

\bibitem[\protect\citeauthoryear{Farrow, Zhu  \& Thrane}{Farrow
  et~al.}{2019}]{Farrow2019}
Farrow N.,  Zhu X.-J.,   Thrane E.,  2019, \mn@doi [The Astrophysical Journal]
  {10.3847/1538-4357/ab12e3}, 876, 18

\bibitem[\protect\citeauthoryear{{Finn}}{{Finn}}{1996}]{Finn1996}
{Finn} L.~S.,  1996, \mn@doi [\prd] {10.1103/PhysRevD.53.2878}, \href
  {http://adsabs.harvard.edu/abs/1996PhRvD..53.2878F} {53, 2878}

\bibitem[\protect\citeauthoryear{{Finn} \& {Chernoff}}{{Finn} \&
  {Chernoff}}{1993}]{Finn1993}
{Finn} L.~S.,  {Chernoff} D.~F.,  1993, \mn@doi [\prd]
  {10.1103/PhysRevD.47.2198}, \href
  {https://ui.adsabs.harvard.edu/abs/1993PhRvD..47.2198F} {47, 2198}

\bibitem[\protect\citeauthoryear{{Hannuksela}, {Haris}, {Ng}, {Kumar}, {Mehta},
  {Keitel}, {Li}  \& {Ajith}}{{Hannuksela} et~al.}{2019}]{Hannuksela2019}
{Hannuksela} O.~A.,  {Haris} K.,  {Ng} K.~K.~Y.,  {Kumar} S.,  {Mehta} A.~K.,
  {Keitel} D.,  {Li} T.~G.~F.,   {Ajith} P.,  2019, \mn@doi [\apjl]
  {10.3847/2041-8213/ab0c0f}, \href
  {https://ui.adsabs.harvard.edu/abs/2019ApJ...874L...2H} {874, L2}

\bibitem[\protect\citeauthoryear{{Haris}, {Mehta}, {Kumar}, {Venumadhav}  \&
  {Ajith}}{{Haris} et~al.}{2018}]{Haris2018}
{Haris} K.,  {Mehta} A.~K.,  {Kumar} S.,  {Venumadhav} T.,   {Ajith} P.,  2018,
  arXiv e-prints, \href {https://ui.adsabs.harvard.edu/abs/2018arXiv180707062H}
  {p. arXiv:1807.07062}

\bibitem[\protect\citeauthoryear{{Hou}, {Fan}  \& {Zhu}}{{Hou}
  et~al.}{2021}]{Hou2021}
{Hou} S.,  {Fan} X.-L.,   {Zhu} Z.-H.,  2021, \mn@doi [\mnras]
  {10.1093/mnras/stab2221}, \href
  {https://ui.adsabs.harvard.edu/abs/2021MNRAS.507..761H} {507, 761}

\bibitem[\protect\citeauthoryear{{Khan}, {Husa}, {Hannam}, {Ohme},
  {P{\"u}rrer}, {Forteza}  \& {Boh{\'e}}}{{Khan} et~al.}{2016}]{Khan2016}
{Khan} S.,  {Husa} S.,  {Hannam} M.,  {Ohme} F.,  {P{\"u}rrer} M.,  {Forteza}
  X.~J.,   {Boh{\'e}} A.,  2016, \mn@doi [\prd] {10.1103/PhysRevD.93.044007},
  \href {https://ui.adsabs.harvard.edu/abs/2016PhRvD..93d4007K} {93, 044007}

\bibitem[\protect\citeauthoryear{{Khan}, {Chatziioannou}, {Hannam}  \&
  {Ohme}}{{Khan} et~al.}{2019}]{Khan2019}
{Khan} S.,  {Chatziioannou} K.,  {Hannam} M.,   {Ohme} F.,  2019, \mn@doi
  [\prd] {10.1103/PhysRevD.100.024059}, \href
  {https://ui.adsabs.harvard.edu/abs/2019PhRvD.100b4059K} {100, 024059}

\bibitem[\protect\citeauthoryear{{LIGO Scientific Collaboration}}{{LIGO
  Scientific Collaboration}}{2018}]{lalsuite}
{LIGO Scientific Collaboration} 2018, {LIGO} {A}lgorithm {L}ibrary -
  {LALS}uite, free software (GPL), \mn@doi{10.7935/GT1W-FZ16}

\bibitem[\protect\citeauthoryear{{Li}, {Mao}, {Zhao}  \& {Lu}}{{Li}
  et~al.}{2018}]{Li2018}
{Li} S.-S.,  {Mao} S.,  {Zhao} Y.,   {Lu} Y.,  2018, \mn@doi [\mnras]
  {10.1093/mnras/sty411}, \href
  {https://ui.adsabs.harvard.edu/abs/2018MNRAS.476.2220L} {476, 2220}

\bibitem[\protect\citeauthoryear{{Liao}, {Fan}, {Ding}, {Biesiada}  \&
  {Zhu}}{{Liao} et~al.}{2017}]{Liao2017}
{Liao} K.,  {Fan} X.-L.,  {Ding} X.,  {Biesiada} M.,   {Zhu} Z.-H.,  2017,
  \mn@doi [Nature Communications] {10.1038/s41467-017-02135-6}, \href
  {https://ui.adsabs.harvard.edu/abs/2017NatCo...8.2136L} {8, 2136}

\bibitem[\protect\citeauthoryear{{Liu}, {Magana Hernandez}  \&
  {Creighton}}{{Liu} et~al.}{2020}]{Liu2020}
{Liu} X.,  {Magana Hernandez} I.,   {Creighton} J.,  2020, arXiv e-prints,
  \href {https://ui.adsabs.harvard.edu/abs/2020arXiv200906539L} {p.
  arXiv:2009.06539}

\bibitem[\protect\citeauthoryear{Madau \& Dickinson}{Madau \&
  Dickinson}{2014}]{Madau2014}
Madau P.,  Dickinson M.,  2014, \mn@doi [Annual Review of Astronomy and
  Astrophysics] {10.1146/annurev-astro-081811-125615}, 52, 415

\bibitem[\protect\citeauthoryear{{Mukherjee}, {Broadhurst}, {Diego}, {Silk}  \&
  {Smoot}}{{Mukherjee} et~al.}{2021a}]{Mukherjee2021-ligo}
{Mukherjee} S.,  {Broadhurst} T.,  {Diego} J.~M.,  {Silk} J.,   {Smoot} G.~F.,
  2021a, \mn@doi [\mnras] {10.1093/mnras/staa3813}, \href
  {https://ui.adsabs.harvard.edu/abs/2021MNRAS.501.2451M} {501, 2451}

\bibitem[\protect\citeauthoryear{{Mukherjee}, {Broadhurst}, {Diego}, {Silk}  \&
  {Smoot}}{{Mukherjee} et~al.}{2021b}]{Mukherjee2021-break}
{Mukherjee} S.,  {Broadhurst} T.,  {Diego} J.~M.,  {Silk} J.,   {Smoot} G.~F.,
  2021b, \mn@doi [\mnras] {10.1093/mnras/stab1980}, \href
  {https://ui.adsabs.harvard.edu/abs/2021MNRAS.506.3751M} {506, 3751}

\bibitem[\protect\citeauthoryear{{Ng}, {Wong}, {Broadhurst}  \& {Li}}{{Ng}
  et~al.}{2018}]{Ng2018}
{Ng} K. K.~Y.,  {Wong} K. W.~K.,  {Broadhurst} T.,   {Li} T. G.~F.,  2018,
  \mn@doi [\prd] {10.1103/PhysRevD.97.023012}, \href
  {https://ui.adsabs.harvard.edu/abs/2018PhRvD..97b3012N} {97, 023012}

\bibitem[\protect\citeauthoryear{{Oguri}}{{Oguri}}{2018}]{Oguri2018}
{Oguri} M.,  2018, \mn@doi [\mnras] {10.1093/mnras/sty2145}, \href
  {https://ui.adsabs.harvard.edu/abs/2018MNRAS.480.3842O} {480, 3842}

\bibitem[\protect\citeauthoryear{{Oguri}}{{Oguri}}{2019}]{Oguri2019}
{Oguri} M.,  2019, \mn@doi [Reports on Progress in Physics]
  {10.1088/1361-6633/ab4fc5}, \href
  {https://ui.adsabs.harvard.edu/abs/2019RPPh...82l6901O} {82, 126901}

\bibitem[\protect\citeauthoryear{P\'erigois, Belczynski, Bulik  \&
  Regimbau}{P\'erigois et~al.}{2021}]{Belczynski2021}
P\'erigois C.,  Belczynski C.,  Bulik T.,   Regimbau T.,  2021, \mn@doi [Phys.
  Rev. D] {10.1103/PhysRevD.103.043002}, 103, 043002

\bibitem[\protect\citeauthoryear{{Pindor}, {Turner}, {Lupton}  \&
  {Brinkmann}}{{Pindor} et~al.}{2003}]{Pindor2003}
{Pindor} B.,  {Turner} E.~L.,  {Lupton} R.~H.,   {Brinkmann} J.,  2003, \mn@doi
  [\aj] {10.1086/374233}, \href
  {https://ui.adsabs.harvard.edu/abs/2003AJ....125.2325P} {125, 2325}

\bibitem[\protect\citeauthoryear{{Pi{\'o}rkowska-Kurpas}, {Hou}, {Biesiada},
  {Ding}, {Cao}, {Fan}, {Kawamura}  \& {Zhu}}{{Pi{\'o}rkowska-Kurpas}
  et~al.}{2020}]{ET5}
{Pi{\'o}rkowska-Kurpas} A.,  {Hou} S.,  {Biesiada} M.,  {Ding} X.,  {Cao} S.,
  {Fan} X.,  {Kawamura} S.,   {Zhu} Z.-H.,  2020, arXiv e-prints, \href
  {https://ui.adsabs.harvard.edu/abs/2020arXiv200508727P} {p. arXiv:2005.08727}

\bibitem[\protect\citeauthoryear{{Pi{\'o}rkowska}, {Biesiada}  \&
  {Zhu}}{{Pi{\'o}rkowska} et~al.}{2013}]{ET1}
{Pi{\'o}rkowska} A.,  {Biesiada} M.,   {Zhu} Z.-H.,  2013, \mn@doi [\jcap]
  {10.1088/1475-7516/2013/10/022}, \href
  {http://adsabs.harvard.edu/abs/2013JCAP...10..022P} {10, 022}

\bibitem[\protect\citeauthoryear{{Robertson}, {Smith}, {Massey}, {Eke},
  {Jauzac}, {Bianconi}  \& {Ryczanowski}}{{Robertson}
  et~al.}{2020}]{Robertson2020}
{Robertson} A.,  {Smith} G.~P.,  {Massey} R.,  {Eke} V.,  {Jauzac} M.,
  {Bianconi} M.,   {Ryczanowski} D.,  2020, \mn@doi [\mnras]
  {10.1093/mnras/staa1429}, \href
  {https://ui.adsabs.harvard.edu/abs/2020MNRAS.495.3727R} {495, 3727}

\bibitem[\protect\citeauthoryear{{Smith} et~al.,}{{Smith}
  et~al.}{2018}]{Smith2018}
{Smith} G.~P.,  et~al., 2018, \mn@doi [IAU Symposium]
  {10.1017/S1743921318003757}, \href
  {https://ui.adsabs.harvard.edu/abs/2018IAUS..338...98S} {338, 98}

\bibitem[\protect\citeauthoryear{{Strolger} et~al.,}{{Strolger}
  et~al.}{2004}]{Strolger2004}
{Strolger} L.-G.,  et~al., 2004, \mn@doi [\apj] {10.1086/422901}, \href
  {https://ui.adsabs.harvard.edu/abs/2004ApJ...613..200S} {613, 200}

\bibitem[\protect\citeauthoryear{Takahashi \& Nakamura}{Takahashi \&
  Nakamura}{2003}]{Takahashi2003}
Takahashi R.,  Nakamura T.,  2003, \mn@doi [The Astrophysical Journal]
  {10.1086/377430}, 595, 1039

\bibitem[\protect\citeauthoryear{{The LIGO Scientific Collaboration}, {The
  Virgo Collaboration}  \& {The KAGRA Scientific Collaboration}}{{The LIGO
  Scientific Collaboration} et~al.}{2021}]{2021LVK}
{The LIGO Scientific Collaboration} {The Virgo Collaboration}  {The KAGRA
  Scientific Collaboration} 2021, arXiv e-prints, \href
  {https://ui.adsabs.harvard.edu/abs/2021arXiv211103634T} {p. arXiv:2111.03634}

\bibitem[\protect\citeauthoryear{{Turner}, {Ostriker}  \& {Gott}}{{Turner}
  et~al.}{1984}]{Turner1984}
{Turner} E.~L.,  {Ostriker} J.~P.,   {Gott} J.~R. I.,  1984, \mn@doi [\apj]
  {10.1086/162379}, \href
  {https://ui.adsabs.harvard.edu/abs/1984ApJ...284....1T} {284, 1}

\bibitem[\protect\citeauthoryear{Wang, Stebbins  \& Turner}{Wang
  et~al.}{1996}]{Wang1996}
Wang Y.,  Stebbins A.,   Turner E.~L.,  1996, \mn@doi [Phys. Rev. Lett.]
  {10.1103/PhysRevLett.77.2875}, 77, 2875

\bibitem[\protect\citeauthoryear{Wang, Lo, Li  \& Chen}{Wang
  et~al.}{2021}]{Wang2021}
Wang Y.,  Lo R. K.~L.,  Li A. K.~Y.,   Chen Y.,  2021, \mn@doi [Phys. Rev. D]
  {10.1103/PhysRevD.103.104055}, 103, 104055

\bibitem[\protect\citeauthoryear{{Wierda}, {Wempe}, {Hannuksela}, {Koopmans}
  \& {Van Den Broeck}}{{Wierda} et~al.}{2021}]{Wierda2021}
{Wierda} A. R. A.~C.,  {Wempe} E.,  {Hannuksela} O.~A.,  {Koopmans} L. V.~E.,
  {Van Den Broeck} C.,  2021, arXiv e-prints, \href
  {https://ui.adsabs.harvard.edu/abs/2021arXiv210606303W} {p. arXiv:2106.06303}

\bibitem[\protect\citeauthoryear{{Xu}, {Ezquiaga}  \& {Holz}}{{Xu}
  et~al.}{2021}]{Xu2021}
{Xu} F.,  {Ezquiaga} J.~M.,   {Holz} D.~E.,  2021, arXiv e-prints, \href
  {https://ui.adsabs.harvard.edu/abs/2021arXiv210514390X} {p. arXiv:2105.14390}

\bibitem[\protect\citeauthoryear{{Yang}, {Ding}, {Biesiada}, {Liao}  \&
  {Zhu}}{{Yang} et~al.}{2019}]{ET4}
{Yang} L.,  {Ding} X.,  {Biesiada} M.,  {Liao} K.,   {Zhu} Z.-H.,  2019,
  \mn@doi [\apj] {10.3847/1538-4357/ab095c}, \href
  {https://ui.adsabs.harvard.edu/abs/2019ApJ...874..139Y} {874, 139}

\bibitem[\protect\citeauthoryear{{Zhu} et~al.,}{{Zhu} et~al.}{2020}]{Zhu2020}
{Zhu} J.-P.,  et~al., 2020, arXiv e-prints, \href
  {https://ui.adsabs.harvard.edu/abs/2020arXiv201102717Z} {p. arXiv:2011.02717}

\makeatother
\end{thebibliography}

\appendix

\section{supplementary prediction}\label{app}
Table \ref{tab:lensed-diffmass} shows predictions of
yearly lensed GW detection rate analogous to Table \ref{tab:lensed}, but with different mass distribution of CBC populations. 
Comparing with Table \ref{tab:lensed}, we note that results for both BNS and NSBH populations do not differ significantly. However, for BBH events modeled with more flat and massive upper-boundary mass distribution, 
the number of lensed BBH events increases to $2.74$ for PlusNetwork and to $>300$ for ET and $>400$ for ET+CE network .

%%%%%%%%%%%%%%%%%%%%%%%%%%%%%%%%%%%%%%%%%%%%%%%%%%

%%%%%%%%%%%%%%%%% APPENDICES %%%%%%%%%%%%%%%%%%%%%

%%%%%%%%%%%%%%%%%%%%%%%%%%%%%%%%%%%%%%%%%%%%%%%%%%
%\appendix
%\section{Selected sample catalogue} 

% Don't change these lines
\bsp	% typesetting comment
\label{lastpage}
\end{document}